\documentclass{elsart}
\usepackage{graphicx}
\begin{document}
\begin{frontmatter}
\title{Optical
transparency of mesoporous metals }
\author[Uni]{N.~Stefanou\thanksref{cauthor}},
\author[Poly]{A.~Modinos} and
\author[Poly]{V.~Yannopapas}
\address[Uni]{University of Athens, Section of Solid State Physics, \\
Panepistimioupolis, GR-157 84 Athens, Greece}
\address[Poly]{Department of Physics, National Technical University
of Athens, \\ Zografou Campus, GR-157 73 Athens, Greece}
\thanks[cauthor]{Corresponding author. Tel. +30-1-7276762;
fax: +30-1-7276711. \\
{\em E-mail address}: nstefan@cc.uoa.gr (N. Stefanou)}

\begin{abstract}
We examine the optical properties of metals containing a periodic
arrangement of nonoverlapping spherical mesopores, empty or
filled with a dielectric material. We show that a slab of such a
porous metal transmits light over regions of frequency determined
by the dielectric constant of the cavities and the fractional
volume occupied by them, with an efficiency which is many orders
of magnitude higher than predicted by standard aperture theory.
Also, the system absorbs light efficiently over the said regions
of frequency unlike the homogeneous metal.



\end{abstract}
\end{frontmatter}

Recently, optical transmission experiments \cite{exp}
have shown that periodic arrays of cylindrical
holes in metallic films display a transmission efficiency,
at wavelengths much larger than the hole diameter, which
is orders of magnitude higher than predicted by standard
aperture theory \cite{bethe}. Such intriguing optical
properties may have important technological applications
in photolithography, in near-field microscopy,
in flat-panel displays, and in novel active filter devices
\cite{exp,techno}. Similar enhanced
optical transmission occurs in metallic gratings
with very narrow slits \cite{pendry},
and it has been recognized that
the physical mechanism behind this phenomenon relies
on resonant transmission through coupled surface-plasmon
modes.

It has long been known that when metals are bombarded with
energetic particles such as neutrons or ions over a sufficiently
long time, regular arrangements of vacancy clusters may form,
such as three-dimensional (3D) lattices of spherical voids or
bubbles. The lattice of voids or bubbles is isomorphic to the
microscopic lattice of the metallic matrix, with a lattice
constant which is typically 100 to 1000 times larger than that of
the matrix lattice (see, e.g., Ref.~ \cite{krishan} and references
therein). Nowadays, considerable advances are being made in the
template-assisted assembly of macroporous and mesoporous metals:
a 3D template is assembled from a self-organizing material and
impregnated with the desired metal; then, the template is removed,
resulting in an array of pores that reflects the structure of the
template \cite{templ}. In this paper we predict some extraordinary
optical transmission effects in such mesoporous metals. We show
that these properties can be understood in a systematic manner on
the basis of exact calculations, which at the end provide a
transparent model of the underlying physics.

We consider, to begin with,
a single spherical cavity, of radius $S$,
in a metal characterized by
a Drude-like relative dielectric function
\begin{equation}
\epsilon=1-{\omega_p^2}/{\omega^2} \;,
\label{drude}
\end{equation}
where $\omega_p$ is the bulk plasma frequency, and we
have neglected damping for now.
The electromagnetic (EM) field at frequency
$\omega$ is described by its electric-field
component
${\bf E}({\bf r};t)=
{\rm Re}[{\bf E}({\bf r}) \exp (-i \omega t)]$,
where ${\bf E}({\bf r})$ is written
as follows (see, e.g., Ref.~ \cite{mod}).
Inside the cavity ($r<S$)
\begin{equation}
{\bf E}({\bf r})=\sum_{l=1}^{\infty} \sum_{m=-l}^{l} \left[ {i
\over {\kappa}_s} a_{l m}^{I E} {\bf \nabla} \times
j_{l}(\kappa_{s}r) {\bf X}_{l m}(\hat{\bf r}) + a_{l m}^{I H}
j_{l}(\kappa_{s}r) {\bf X}_{l m} (\hat{\bf r}) \right]  \;,
\label{field:in}
\end{equation}
where $\kappa_{s}=\omega /c$, $c$ being the velocity of light
in vacuum; $j_{l}$ is a spherical Bessel function; and
${\bf X}_{l m}(\hat{\bf r})$ is
a vector spherical harmonic.
We need not write down the associated
magnetic-field component of the EM wave. Outside the cavity
($r>S$)

\begin{eqnarray}
{\bf E}({\bf r})=\sum_{l=1}^{\infty} \sum_{m=-l}^{l} \biggl[  {i
\over \kappa} a_{l m}^{0 E} {\bf \nabla} \times j_{l}(\kappa r)
{\bf X}_{l m}(\hat{\bf r}) &+& a_{l m}^{0 H} j_{l}(\kappa r) {\bf
X}_{l m}(\hat{\bf r})
 \nonumber \\
 +
{i \over {\kappa}} a_{l m}^{+ E} {\bf \nabla} \times
h_{l}^{+}(\kappa r) {\bf X}_{l m}(\hat{\bf r}) & + &  a_{l m}^{+
H} h_{l}^{+}(\kappa r) {\bf X}_{l m} (\hat{\bf r}) \biggr] \;,
\label{field:out}
\end{eqnarray}
where $h_{l}^{+}$ is a spherical Hankel function
corresponding to an outgoing wave.
The first two terms in the above equation describe an
incident wave and the last two terms a scattered wave.
The wavenumber $\kappa$ for a medium with
negative dielectric function, as is the case to
be considered here ($\omega < \omega_{p}$),
is a purely imaginary number:
$\kappa=iq=i\omega \sqrt{-\epsilon} /c$.

Because of the spherical symmetry of the scatterer we obtain
\begin{eqnarray}
a_{l m}^{+ E}&=&T_{l}^{E} a_{l m}^{0 E}\nonumber \\
a_{l m}^{+ H}&=&T_{l}^{H} a_{l m}^{0 H} \;, \label{scat}
\end{eqnarray}
where
\begin{equation}
T_{l}^{E}(\omega)={\left[
{j_{l}({\kappa}_{s}r) {\partial  \over \partial r}
(rj_{l}({\kappa}r)) }-
 j_{l}({\kappa}r) {\partial  \over \partial r}
(rj_{l}({\kappa}_{s}r)) {\epsilon}
 \over
 h_{l}^{+}({\kappa}r) {\partial  \over \partial r}
(rj_{l}({\kappa}_{s}r)) {\epsilon}-
 j_{l}({\kappa}_{s}r) {\partial  \over \partial r}
(rh_{l}^{+}({\kappa}r))
\right]}_{r=S}
\label{tmat}
\end{equation}
with a corresponding expression for $T_{l}^{H}(\omega)$
\cite{mod}. In the case of a single cavity in a homogeneous
medium of
negative dielectric function there can be no incident wave:
$a_{l m}^{0 E}=a_{l m}^{0 H}=0$
and, therefore, non-trivial states of the EM field, of given
$l$, will exist if
\begin{equation}
T_{l}^{E}(\omega)=\infty
\ \ \ \   \mbox{or} \ \ \ \
T_{l}^{H}(\omega)=\infty \;.
\label{cond1}
\end{equation}
It can be shown that
the second of the above equations does not obtain
a real frequency solution in the region
from $0$ to $\omega_{p}$.
The first equation, for any $l$, has a solution
in this frequency region; we denote it by
$\tilde \omega_l$.
For a small cavity, using the appropriate asymptotic
expansions of the spherical Bessel and Hankel functions for
small arguments \cite{arfken} in Eq.~(\ref{tmat}),
we obtain
\begin{equation}
\tilde \omega_{l} \simeq \omega_{p} \sqrt{(l+1)/(2l+1)} \;,
\ \ \ \ l=1,2,3,... \;.
\label{plfre}
\end{equation}
Taking for instance $S \omega_{p} / c =0.4$,
which for $\omega_{p}=10$ eV corresponds to a radius
$S=7.5$ nm, Eq.~(\ref{plfre}) approximates $\tilde \omega_l$
with an accuracy better than $1 \%$. The eigenmodes of the
EM field obtained at $\tilde \omega_l$ define the
$2^l$-pole plasma oscillations at the surface of the cavity.

Let us now ask whether it is possible for a system of small
spherical cavities in a metal to be transparent.
For a crude, qualitative description of the optical properties
of the composite system, we can use its
effective dielectric constant, $\overline \epsilon$,
as given by the Maxwell Garnett (MG)
effective-medium theory \cite{mg}
\begin{equation}
\frac {\overline \epsilon - \epsilon}
{\overline \epsilon +2\epsilon} = f \frac
{\epsilon_{s} - \epsilon}{\epsilon_{s} +2\epsilon} \;,
\label{mg}
\end{equation}
where $f$ is the fractional volume occupied by the cavities
($\epsilon_{s}=1$) and $\epsilon$ is the relative dielectric
function of the metal given by Eq.~(\ref{drude}). Eq.~(\ref{mg})
gives a positive $\overline \epsilon$ over the range of
frequencies from $\omega_{\rm min}=\omega_{p} \sqrt{2(1-f)/3}$ to
$\omega_{\rm max}=\omega_{p} \sqrt{(2+f)/3}$. In other words, in
the above frequency region, there is a band of propagating states
of the EM field in the porous metal, the width of which,
$\omega_{\rm max}-\omega_{\rm min}$, increases with $f$. As $f
\rightarrow 0$, $\omega_{max} \rightarrow \omega_{min} \rightarrow
\omega_{p} \sqrt{2/3}$ which, according to Eq.~(\ref{plfre}), is
the resonance frequency, $\tilde \omega_{1}$, associated with the
dipole plasma oscillations at the surface of a single small
cavity. Clearly the above band (for $f>0$) arises from the
interaction between dipole plasma modes localized on neighboring
cavities and can therefore be understood in the spirit of the
tight-binding approximation \cite{stef98}. Its bandwidth will be
larger the larger the spatial overlap between the wavefields
associated with neighboring spheres which in turn increases with
$f$.

The entire band can be shifted to lower frequencies by
filling the pores with a dielectric material.
One can show that the resonance frequencies
of a single small sphere of dielectric constant $\epsilon_s$
in a metal described by Eq.~(\ref{drude}) are given by
\begin{equation}
\tilde \omega_{l} \simeq \omega_{p}
\sqrt{(l+1)/(l \epsilon_{s}+l+1)} \;.
\label{resdie}
\end{equation}
Therefore, with $\epsilon_{s} \simeq 10$, $\tilde \omega_1$
of the isolated sphere and the corresponding
band of the porous metal are found in the optical region.

We note that the MG theory is based on the
electric-dipole approximation and, therefore, cannot describe
$2^l$-pole states with $l>1$.
There is of course an infinite number of bands
corresponding to the
$2^l$-pole plasma modes of the individual spheres.
These modes, according to Eq.~(\ref{resdie}), extend
approximately from
$\omega_{p} \sqrt{1/(\epsilon_{s} +1)}$ to
$\omega_{p} \sqrt{2/(\epsilon_{s} +2)}$.
It is therefore clear that there are no frequency bands
below a cutoff frequency at about
$\omega_{p} \sqrt{1/(\epsilon_{s} +1)}$.
This is what one expects of a metal (a system with DC
conductivity at any temperature including zero), and the
system under consideration possesses DC conductivity
because the metallic component of it forms a
continuous network \cite{topo}. It is understood, however,
that at low frequencies ($\omega \tau < 1$) Eq.~(\ref{drude})
must be replaced by Eq.~(\ref{drude:damp}) below,
which takes into
account the damping of conduction-band electrons.
We note that, in a system consisting of
nonoverlapping metallic spheres in a dielectric medium,
there are
propagating modes of the EM field at low frequencies
with a free-photon-like dispersion \cite{yanno99}, reflecting
the fact that the system is an insulator at $\omega = 0$.

An accurate analysis of the optical properties of porous metals
can be efficiently carried out using the method we developed for
the calculation of the optical properties of photonic crystals
consisting of nonoverlapping spheres in a homogeneous host medium.
The details of the method and a computer program for its
implementation can be found elsewhere \cite{stef92,comphy}. Here
we need only say that in the present case, of a metallic host
medium, where only evanescent EM waves exist below the plasma
frequency, the ${\bf \Omega}$-matrices (as defined in
Ref.~\cite{comphy}) must be calculated by a direct summation over
the space lattice; a few terms in the relevant lattice sums give
adequate convergence.

We consider a system of nonoverlapping, identical silicon spheres
($\epsilon_{s}=11.9$), centered at the sites of a fcc lattice of
lattice constant $a$, in a metallic host medium whose dielectric
function is given by Eq.~(\ref{drude}). The radius, $S$, of the
spheres equals one fifth of the first-neighbor distance, $a_0$,
which corresponds to a fractional volume occupied by the spheres
$f=4.74 \%$. We view the crystal as a stack of layers (planes of
spheres) parallel to the fcc (001) surface. For given ${\bf
k}_{\parallel}$, the reduced component of the wavevector parallel
to the fcc (001) surface, we calculate, as functions of $\omega$,
the frequency lines $k_{z}=k_{z}(\omega ; {\bf k}_{\parallel})$
corresponding to generalized Bloch wave solutions of the EM field.
$k_{z}(\omega ; {\bf k}_{\parallel})$ is the $z$ component [normal
to the (001) plane] of the wave vector of a generalized Bloch wave
with the given $\omega$ and ${\bf k}_{\parallel}$; and there are
many such waves corresponding to different values of $k_z$. The
regions of $\omega$ over which at least one $k_z$ is real define
corresponding frequency bands and regions over which all $k_z$ are
complex define frequency gaps, for the given ${\bf
k}_{\parallel}$.

\begin{figure}
\centerline{\includegraphics*[width=12.0cm]{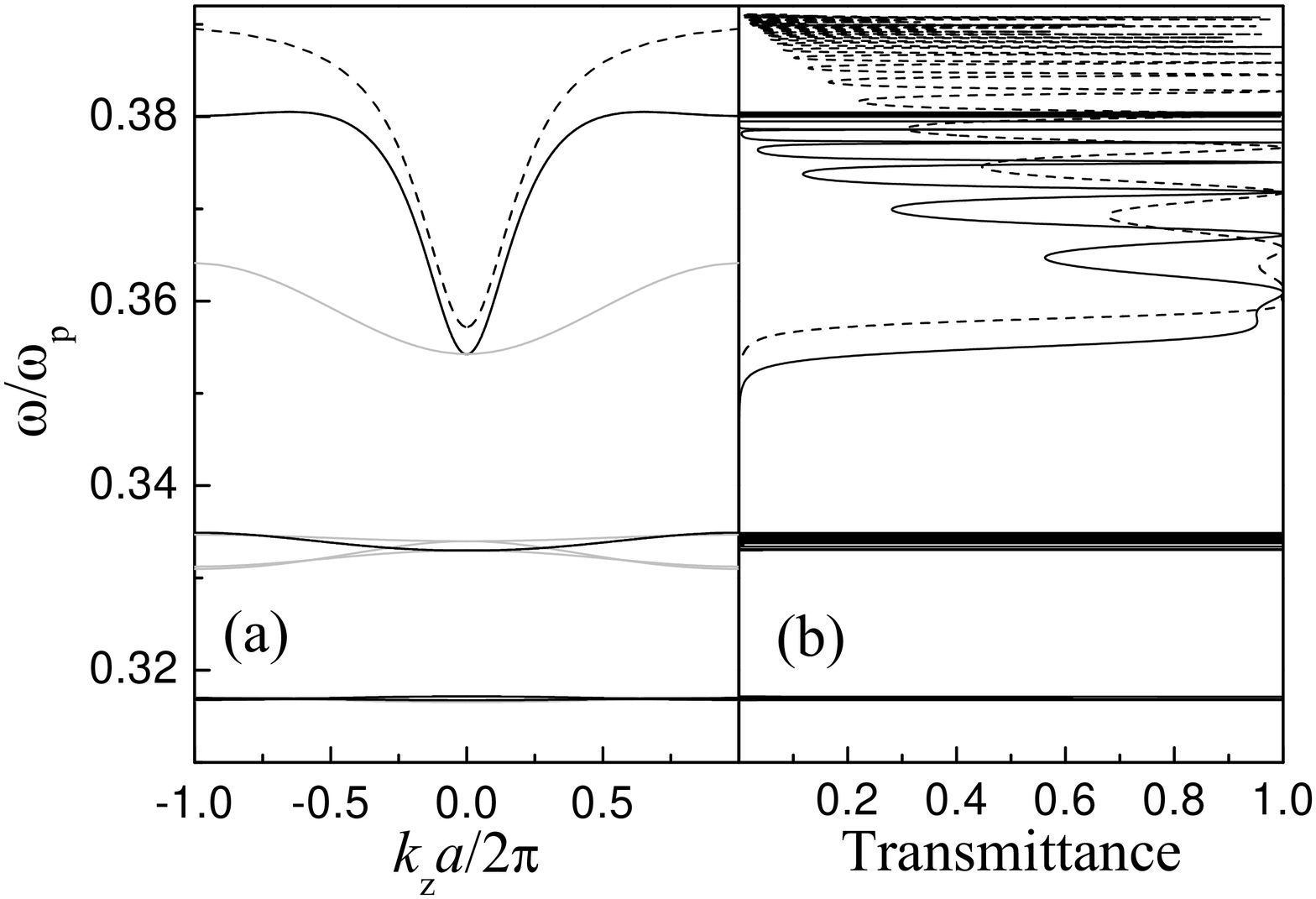}}
\caption{
(a) The photonic band structure normal to the (001) surface of a
fcc crystal of silicon spheres in a nonabsorbing Drude metal
($a_{0} \omega_{p} /c =1, \; S \omega_{p} /c =0.2$). The solid
lines (black lines: doubly degenerate bands, gray lines:
nondegenerate bands) are exact results; the broken line is the MG
result. (b) The corresponding transmittance curve for light
incident normally on a slab of the above crystal consisting of 16
lattice planes parallel to the (001) surface. The solid (broken)
lines are exact (MG) results. } \label{fig1}
\end{figure}

\begin{figure}
\centerline{\includegraphics*[width=12.0cm]{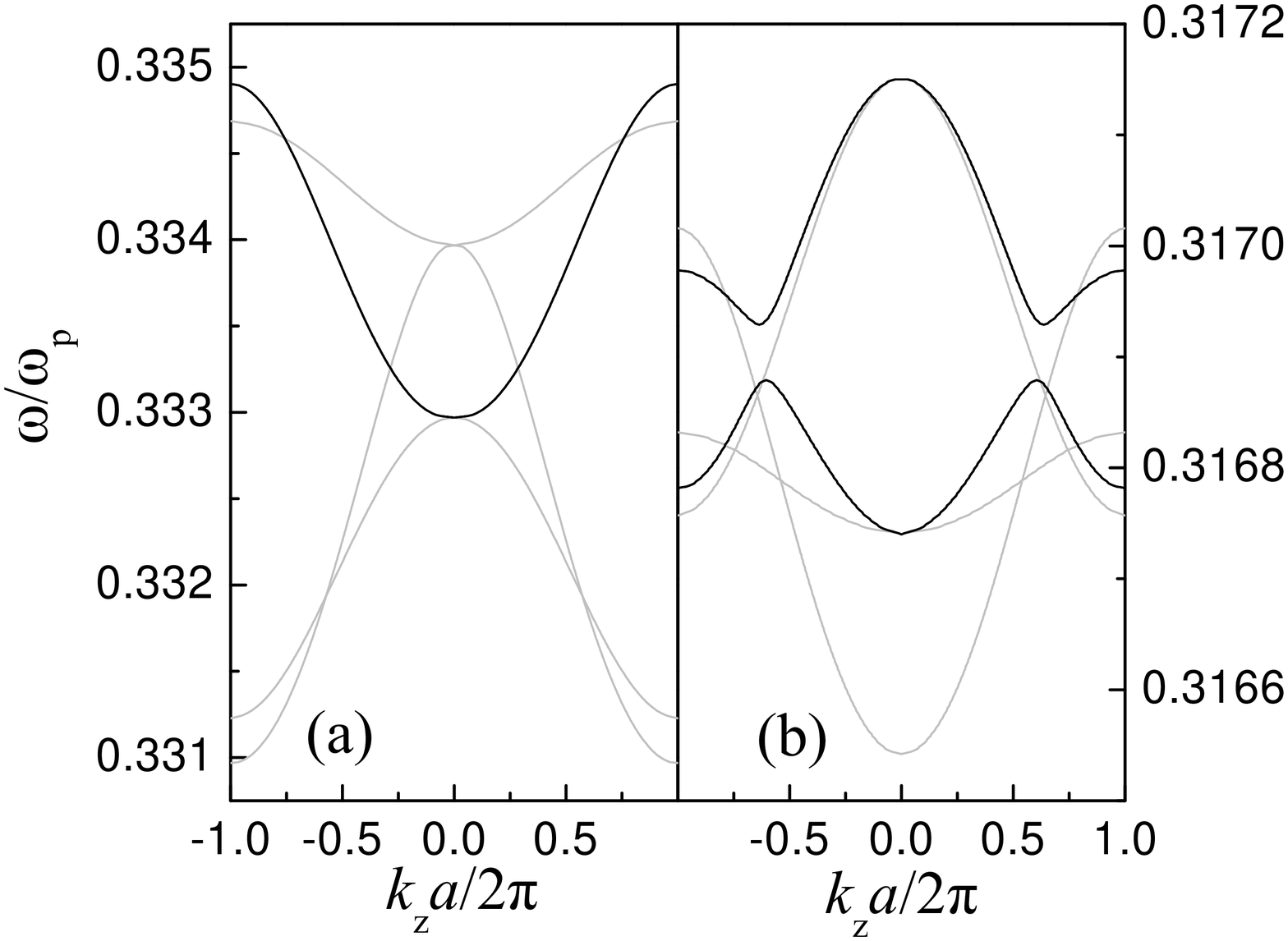}}
 \caption{ The frequency
bands associated with the $l=2$ (a) and $l=3$ (b) states of an
isolated sphere, for the system described in Fig.~\ref{fig1}. }
\label{fig2}
\end{figure}

In Fig.~\ref{fig1}a we show the frequency bands for ${\bf
k}_{\parallel}={\bf 0}$ [dispersion curves along the normal to the
(001) plane] of the above crystal, which arise from the $2^l$-pole
states of the individual spheres for $l=1,2,3$. We see that these
bands develop about the corresponding resonance frequencies given
by Eq.~(\ref{resdie}), and that their width decreases with $l$.
Apparently the spatial extent of the wavefield around a sphere
decreases with increasing $l$, thus leading to a weaker
hybridization between the higher $2^l$-pole states of the spheres
and, consequently, to a smaller bandwidth. The bands shown by the
black solid lines in Fig.~\ref{fig1}a are doubly degenerate and
couple with light incident normally on a slab of the crystal
parallel to the (001) surface. The black broken line shows the
dipole band (doubly degenerate) as obtained in the MG
approximation. The bands shown by the gray solid lines are
nondegenerate and do not couple with normally incident light. The
existence of these optically inactive modes at high-symmetry
points and along symmetry lines has been noted by a number of
authors \cite{stef92,inactive}. Fig.~\ref{fig2} shows more
clearly all frequency bands corresponding to $l=2$ and $l=3$. One
can see that the total number of bands associated with a given
$l$ equals $2l+1$ as expected from the $(2l+1)$-degeneracy of the
corresponding state of the isolated sphere. The degeneracy of the
bands and the symmetry properties of the corresponding eigenmodes
is what one expects from a group-theoretical analysis
\cite{group}. Next to the frequency band structure, in
Fig.~\ref{fig1}b, we show by the solid line the transmission
coefficient of light incident normally on a slab of the crystal
consisting of 16 planes of spheres parallel to the (001) surface,
assuming that the medium on either side of the slab is air. As
expected, the transmittance practically vanishes for frequencies
within the frequency gaps of the infinite crystal and exhibits the
well-known Fabry-P\'erot-type resonances over the regions of the
optically active frequency bands \cite{yanno99}. The broken line
shows the transmittance as evaluated in the MG approximation.

\begin{figure}
\centerline{\includegraphics*[width=12.0cm]{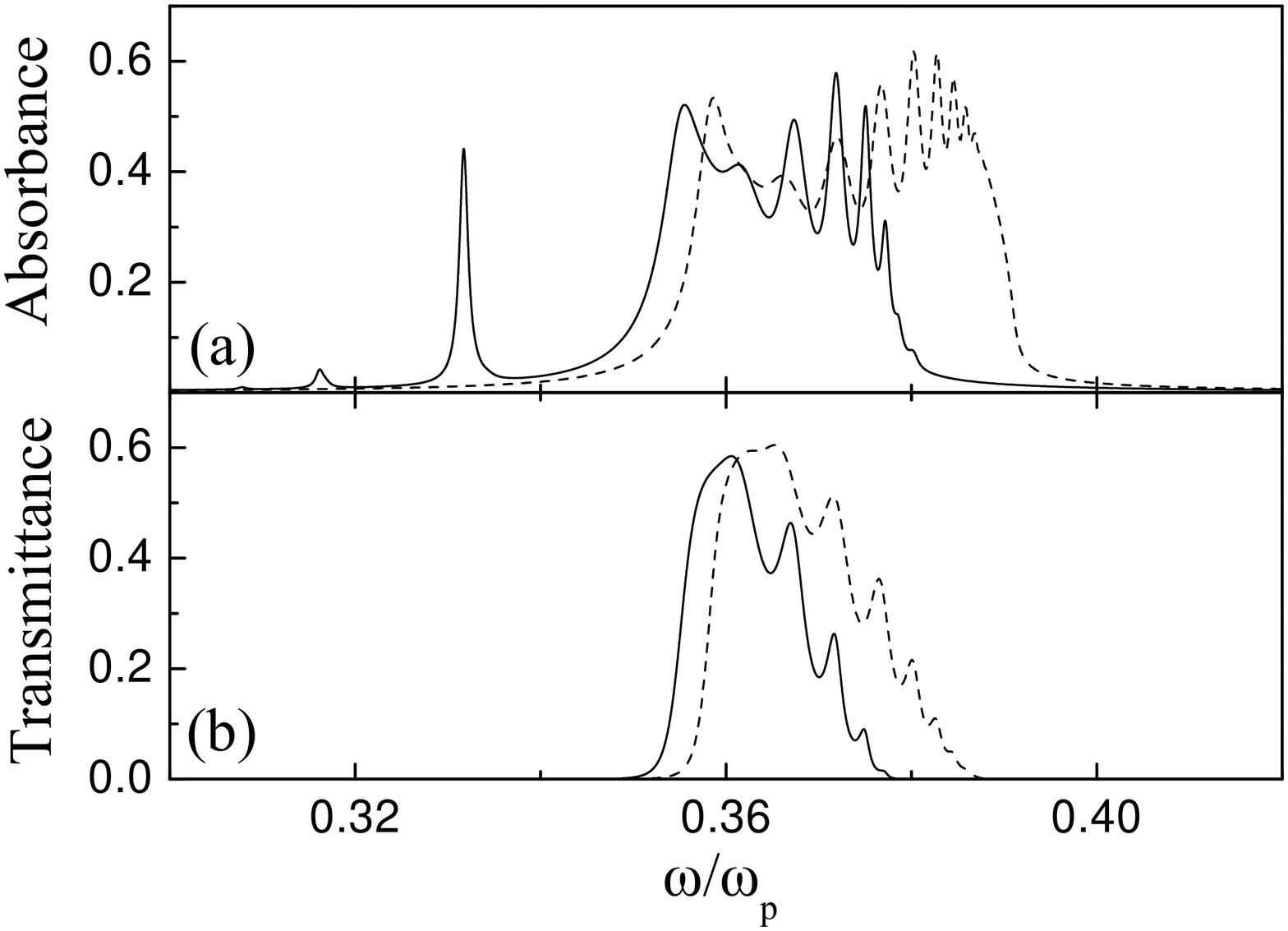}}
 \caption{ Absorbance (a) and
transmittance (b) of light incident normally on a slab of 16
lattice planes parallel to the (001) surface of a fcc crystal
consisting of silicon spheres in a Drude metal [$a_{0} \omega_{p}
/c =1, \; S \omega_{p} /c =0.2, \; (\omega_{p} \tau)^{-1}=0.001$].
The solid (broken) lines are exact (MG) results. } \label{fig3}
\end{figure}

The absorption of light by the material, which we have
neglected so far, is of course
a quantity of great importance. We account for it by
inserting, as usual, a damping term in the relative
dielectric
function given by Eq.~(\ref{drude}). We write
\begin{equation}
\epsilon=1-\frac{\omega_p^2}
{\omega(\omega+i \tau^{-1})} \;,
\label{drude:damp}
\end{equation}
where $\tau$ is the relaxation time of the conduction-band
electrons, with a typical value
$(\omega_{p} \tau)^{-1}=0.001$.
Absorption is mainly associated with the
resonantly oscillating multipoles of the spheres and,
therefore, is appreciable only in the region of the
frequency bands. This is demonstrated
in Fig.~\ref{fig3}a, for light incident normally on a slab
of the crystal consisting of 16 planes of spheres
parallel to the (001) surface. We see that absorption is
considerable; the
corresponding absorbance of a homogeneous metallic slab
of the same thickness is of the order of $10^{-3}$.
In Fig.~\ref{fig3}b we show the transmission coefficient
of light for the same system; transmission
occurs in the frequency region of the dipole band.
It is somewhat reduced, due to absorption, but it is
extraordinarily high: the corresponding transmittance
of a homogeneous metallic slab of the same thickness is
of the order of $10^{-9}$, while a single hole of
diameter $d$ in a metal film transmits light of
wavelengths $\lambda >> d$ with an efficiency
proportional to $(d/ \lambda)^4$ \cite{bethe} and in
our case we have $d/ \lambda \approx 0.02$.

In addition to the above, the results shown in Figs.~\ref{fig1}
and \ref{fig3} indicate that though the MG approximation gives a
reasonable estimate of the transmittance and absorbance of the
system in the frequency region of the dipole band, it differs
significantly from the exact picture. And of course, it fails to
reproduce all features associated with the higher $2^l$-pole
resonances (see also Ref.~\cite{yanno99}). Finally, it appears
that the position of the frequency bands (and consequently the
range of frequencies over which a high transmission and
absorption occurs) depends mostly on the characteristics of a
single sphere in the metal host and on the fractional volume
occupied by the spheres, and less so on the specific arrangement
of the spheres. We should also emphasize that though our results
have been obtained for a metallic component described by
Eqs.~(\ref{drude}) and (\ref{drude:damp}), what really matters is
that the dielectric function of the host medium is such that
solutions of Eqs.~(\ref{cond1}) exist at appropriate frequencies;
and as long as this remains true, similar results will be
obtained.

\section*{Acknowledgements}
 We thank Dr. I.~E.~Psarobas for useful discussions. V.
Yannopapas is supported by the State Foundation (I.K.Y.) of
Greece.

\newpage


\begin{thebibliography}{9}
\bibitem{exp} T.~W.~Ebbesen, H.~J.~Lezec, H.~F.~Ghaemi,
T.~Thio, and P.~A.~Wolff, Nature (London) 391 (1998) 667;
H.~F.~Ghaemi, T.~Thio, D.~E.~Grupp, T.~W.~Ebbesen, and H.~J.~
Lezec, Phys. Rev. B 58 (1998) 6779; T.~J.~Kim, T.~Thio,
T.~W.~Ebbesen, D.~E.~Grupp, and H.~J.~Lezec, Opt. Lett. 24 (1999)
256.
\bibitem{bethe} H.~A.~Bethe, Phys. Rev. 66 (1944) 163.
\bibitem{techno} J.~R.~Samples, Nature (London) 391 (1998)
641; P.~R.~Villeneuve, Phys. World 11 (1998) 28.
\bibitem{pendry} J.~B.~Pendry, Science 285 (1999) 1687;
J.~A.~Porto, F.~J.~Garc\'{\i}a-Vidal, and J.~B.~Pendry, Phys.
Rev. Lett. 83 (1999) 2845.
\bibitem{krishan} K.~Krishan, Radiation Effects 66 (1982) 121.
\bibitem{templ} P.~Jiang, J.~Cizeron, J.~F.~Bertone, and
V.~L.~Colvin, J. Am. Chem. Soc. 121 (1999) 7957; O.~D.~Velev,
P.~M.~Tessier, A.~M.~Lenhoff, and E.~W.~Kaler, Nature (London)
401 (1999) 548; H.~Yan, C.~F.~Blanford, B.~T.~Holland, M.~Parent,
W.~H.~Smyrl, and A.~Stein, Adv. Mater. 11 (1999) 1003;
O.~D.~Velev and E.~W.~Kaler, Adv. Mater. 12 (2000) 531;
K.~M.~Kulinowski, P.~Jiang, H.~Vaswani, and V.~L.~Colvin, Adv.
Mater. 12 (2000) 833; J.~E.~G.~J.~Wijnhoven,
S.~J.~M.~Zevenhuizen, M.~A.~Hendriks, D.~Vanmaekelbergh,
J.~J.~Kelly, and W.~L.~Vos, Adv. Mater. 12 (2000) 888; N.~Eradat,
J.~D.~Huang, Z.~V.~Vardeny, A.~A.~Zakhidov, and R.~H.~Baughman
(to be published).
\bibitem{mod} A.~Modinos, Physica 141A (1987) 575.
\bibitem{arfken} G.~Arfken, Mathematical Methods for
Physicists, Academic Press, New York, 1970.
\bibitem{mg} J.~C.~Maxwell Garnett, Philos. Trans. R. Soc.
London, Ser. A 203 (1904) 385; 205 (1906) 237.
\bibitem{stef98} N.~Stefanou and A.~Modinos, Phys. Rev. B
57 (1998) 12127.
\bibitem{topo} M.~M.~Sigalas, C.~T.~Chan, K.~M.~Ho, and
C.~M.~Soukoulis, Phys. Rev. B 52 (1995) 11744; D.~F.~Sievenpiper,
M.~E.~Sickmiller, and E.~Yablonovitch, Phys. Rev. Lett. 76 (1996)
2480; J.~B.~Pendry, A.~J.~Holden, W.~J.~Stewart, and I.~Youngs,
Phys. Rev. Lett. 76 (1996) 4773; J.~B.~Pendry, A.~J.~Holden,
D.~J.~Robins, and W.~J.~Stewart, J. Phys.: Condens. Matter 10
(1998) 4785.
\bibitem{yanno99} V.~Yannopapas, A.~Modinos, and N.~Stefanou,
Phys. Rev. B 60 (1999) 5359.
\bibitem{stef92} N.~Stefanou, V.~Karathanos, and A.~Modinos,
J. Phys.: Condens, Matter 4 (1992) 7389.
\bibitem{comphy} N.~Stefanou, V.~Yannopapas, and A.~Modinos,
Comput. Phys. Commun. 113 (1998) 49; 132 (2000) 189.
\bibitem{inactive} W.~M.~Robertson, G.~Arjavalingam, R.~D.~
Meade, K.~D.~Brommer, A.~M.~Rappe, and J.~D.~Joannopoulos, Phys.
Rev. Lett. 68 (1992) 2023; K.~Ohtaka and Y.~Tanabe, J. Phys. Soc.
Jpn. 65 (1996) 2670; V.~Karathanos, J. Mod. Opt. 45 (1998) 1751.
\bibitem{group} K.~Ohtaka and Y.~Tanabe, J. Phys. Soc. Jpn.
65 (1996) 2670.
\end{thebibliography}
\end{document}